# Phasing Parameter Analysis for Satellite Collision Avoidance in Starlink and Kuiper Constellations

Jintao Liang, Aizaz U. Chaudhry, and Halim Yanikomeroglu
Department of Systems and Computer Engineering, Carleton University, Ottawa, Canada–K1S 5B6
{jintaoliang}@cmail.carleton.ca, {auhchaud, halim}@sce.carleton.ca

*Abstract*—The phasing parameter *F* determines the relative phasing between satellites in different orbital planes and thereby affects the relative position of the satellites in a constellation. The collisions between satellites within the constellation can be avoided if the minimum distance among them is large. From among the possible values of *F* in a constellation, a value of *F* is desired that leads to the maximum value of the minimum distance between satellites. We investigate *F* for two biggest upcoming satellite constellations including Starlink Phase 1 Version 3 and Kuiper Shell 2. No existing work or FCC filing provides a value of *F* that is suitable for these two constellations. We look for the best value of *F* in these constellations that provides the maximum value of the minimum distance to ensure intra-constellation avoidance of collisions between satellites. To this end, we simulate each constellation for each value of *F* to find its best value based on ranking. Out of the 22 and 36 possible values of *F* for Starlink Phase 1 Version 3 and Kuiper Shell 2, respectively, it is observed that the best value of *F* with highest ranking is 17 and 11 that leads to the largest minimum distance between satellites of 61.83 km and 55.89 km in these constellations, respectively.

## I. INTRODUCTION

Satellites have a long history, and with the development of the technologies for satellite and wireless communication, they are now used for many applications, including wireless communication, vehicle navigation, and terrestrial imaging [1]. Satellite communication has many advantages compared to terrestrial communication, like larger coverage [2] and smaller propagation delay over long distances [3][4]. The requirements for data latency, accuracy, and real-time observations are getting higher and stricter, and in order to satisfy these demands, complex satellite constellations with large number of satellites need to be used and their performance has to be analyzed.

Generally, a Walker constellation pattern is used to design a satellite constellation. Based on their patterns, the Walker constellations can be divided into two categories: Walker Star constellations and Walker Delta constellations. The Walker Star pattern provides a polar constellation, the inclination of the constellation is around 90°, and the orbital planes cross the Poles. The Walker Delta pattern produces an inclined constellation, the inclination is smaller than 90° and is usually between 30° and 60°, and the orbital planes don't cross the Poles and yet cover most of the Earth's surface with high population density. The Walker constellations use a notation of *i*: *T/P/F* [5], where a total number of *T* satellites is evenly distributed in *P* orbital planes, and the inclination is *i* degrees. The phasing parameter *F* can take integer values from 0 to *P*-1 and affects the relative phasing $\beta$ among satellites in different orbital planes.

The single largest threat to a complex satellite constellation is the collision between satellites, which can result in the destruction of satellites and can also create debris harmful to other satellites. Based on the nature of satellites' encounter in space, the collisions between them can be classified into two categories: intra-constellation collisions that can happen between satellites within the same constellation; and inter-constellation collisions that can occur between satellites in different constellations. To design a constellation for collision avoidance within the same constellation, phasing parameter *F* is an important factor that needs to be determined since it affects the minimum distance between satellites. If this distance is smaller than a certain minimum value based on the size of satellites within a constellation, there can be collision between two orbiting satellites. There are many mature controls and techniques to avoid collision when the satellites come close to each other yet finding an optimal value of *F* is the first step to collision avoidance based on the fundamental design of the constellation.

In this work, we analyze *F* in two big and upcoming satellite constellations, namely Starlink Phase 1 Version 3 [6] and Kuiper Shell 2 [7]. Starlink's Phase 1 constellation has changed multiple versions since its initial design. We are interested in its latest version, i.e., Version 3. Both Starlink Phase 1 Version 3 and Kuiper Shell 2 constellations are on low Earth orbit and inclined, and also have more than 1,000 satellites. Such complicated constellations require more strict and suitable choice of *F* to guarantee the safety of the satellites since *F* determines the relative phasing between satellites, and thereby affects the position of satellites in the constellation and the minimum distance between them.

While designing a constellation, the best value of *F* is desired that provides the largest minimum distance between satellites for intra-constellation satellite collision avoidance. To find the best value of *F* in Starlink Phase 1 Version 3 and Kuiper Shell 2 constellations, we simulate each of them for all possible values of *F* and study the effect of *F* on the minimum distance between satellites. For each constellation, the values of *F* range from 0 to *P*-1 and we rank them based on minimum distance between satellites. For Starlink Phase 1 Version 3, the best value of *F* is found to be 17 that offers the largest minimum distance between satellites of 61.83 km while for Kuiper Shell 2, the best value is 11 that gives the largest minimum distance of 55.89 km. No existing work or FCC filing provides the phasing parameter information about Starlink Phase 1 Version 3 and Kuiper Shell 2 constellations. There has been a study on *F* for Version 1 of Starlink's Phase 1 constellation [8], however, in this work we focus on the latest version of this constellation, i.e., Version 3. To the best of our knowledge, this work is the first study to investigate the best value of *F* for Starlink Phase 1 Version 3 and Kuiper Shell 2 constellations.

The rest of the paper is organized as follows. The related work is discussed in Section II. Section III presents a brief introduction of the two constellations, namely Starlink Phase 1 Version 3 and Kuiper Shell 2. The concepts of phasing parameter and collision avoidance, and the methodology of the phasing parameter analysis are given in Section IV. Section V provides the results of the phasing parameter analysis for the two constellations. Conclusions are summarized in Section VI.

## II. LITERATURE REVIEW

In space, if a satellite constellation has a number of orbital planes while each has equal number of satellites within it, then the constellation can be regarded as a uniform grid around the globe, and each satellite is a vertex of this grid [9]. The Walker constellation pattern then can be considered to design and construct the constellation which is uniform. The Walker approach uses symmetric distribution of satellites on orbital planes. For inclined constellations, satellites are closer when they are at higher latitudes near the Poles and sparser when near the Equator.

In the process of designing a constellation, phasing parameter is one of the important parameters to determine. Since orbital perturbation, including turbulence and collision between satellites, threatens the safety and stability of a satellite constellation, the relative position and distance between satellites should be monitored and kept at safety limits [10]. In this way, studying phasing parameter is very important to find the best relative phasing between satellites that leads to the maximum value of the minimum distance between them for intra-constellation satellite collision avoidance.

There are many previous studies that focus on constellation design, including the analysis of the phasing parameter, yet these are mainly based on old constellations like Iridium. In [11], the selected value of $F$ for Galileo Navigation System is found to be 1, which is a Walker Delta 56°:27/3/1 constellation. In [12], the practical value of $F$ for Iridium constellation is taken as 2, while the Walker Star notation for this constellation is 86.4°:66/6/2. In [8], the constellation studied for Version 1 of Starlink's Phase 1 is at an altitude of 1,150 km and has a notation of 53°:1,600/32/$F$. According to the analysis of the minimum passing distance between satellites in this study, the most suitable value of $F$ is 5 that provides the maximum value of the minimum distance to minimize the probability of collision. For different constellations the best value of $F$ varies, yet its value is always taken when $\beta$ is most ideal, i.e., when it provides the largest minimum distance between satellites.

If satellites are involved in a collision, debris and pieces generated as a result will be a catastrophe to other satellites. A satellite collision happened in 2009 when Iridium 33 and Cosmos 2251 satellites collided at 800 km altitude creating more than 2,000 pieces of debris with diameters larger than 10 cm and thousands of smaller pieces [13]. Much of this debris will remain in orbit for quite a long period posing a collision risk to other satellites in LEO satellite constellations. The worst-case scenario resulting from a satellite collision is described by the Kessler Syndrome, which is a theory proposed by NASA scientist Donald J. Kessler in 1978. It is used to describe a cascading and developing phenomenon of collisions caused by the debris in space. He believed that once a certain collision happens, the total amount of space debris will keep on increasing since one collision can create debris which can lead to more collisions in the form of a chain reaction, and this will lead to an exponential growth in collisions producing a belt of debris around the Earth [14].

Big space institutions like ESA (European Space Agency) are not just seeking approaches to avoid debris production in space but are looking to find ways to reduce the total mass of current debris as well. There is urgent need for the cleaning of debris in space, since debris levels have increased 50% in the last five years in low Earth orbit. Thus, the removal of large-sized satellite remnants and debris (caused from previous collisions) from satellite flying orbits to disposal orbits is vital for the safety of existing and upcoming satellite constellations. A debris removal system is introduced in [15], which uses a conductive disposable electro-dynamic tether for the capture of debris. It proposes a debris removal satellite to remove the debris objects in space by capturing and transferring them to lower disposal orbits by robotic arm and tether. Once at lower disposal orbits, the debris will eventually re-enter Earth's atmosphere and is removed during burn up.

## III. SATELLITE CONSTELLATIONS

This section introduces Starlink Phase 1 Version 3 and Kuiper Shell 2 constellations, as described in their FCC filings and press releases as of 2019.

### A. Starlink Phase 1 Version 3 Constellation

SpaceX has a tremendous and promising future blueprint of its Starlink mega-constellation, which comprises 4,408 satellites that will be distributed across five LEO orbits or shells. Starlink will use the Ku band for user communications, and gateway communications will be carried out in Ka band. It will provide satellite Internet access worldwide, and SpaceX plans to sell some of the satellites for certain military, scientific, or exploratory purposes.

In this work, we study the Phase 1 constellation of the Starlink mega-constellation. SpaceX has already deployed hundreds of satellites as part of the Phase 1 of Starlink and this will be the first constellation to become operational among Starlink constellations. The latest version of Starlink's Phase 1 constellation contains 1,584 satellites and has 22 orbital planes spaced 16.4° apart with 72 satellites in each plane. Its altitude is 550 km and its inclination is 53°. Its Walker constellation notation is 53°:1,584/22/$F$, with $F$ as the phasing parameter whose value is not known. Fig. 1 shows the constellation pattern of Starlink Phase 1 Version 3 constellation generated when $F$ is equal to 17.

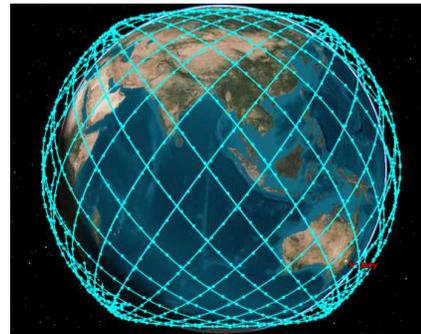

**Fig. 1.** Starlink Phase 1 Version 3 constellation.

### B. Kuiper Shell 2 Constellation

Kuiper satellite constellation has been designed by Amazon, and it aims to provide broadband Internet connectivity across the United States. Amazon announced that it will cost US$10 billion in designing and deploying this satellite constellation. The Kuiper constellation is planned to consist of 3,236 satellites operating in 98 orbital planes in three different orbital shells, one each at 590 km, 610 km, and 630 km altitude.

In this work, we focus on the second shell of the Kuiper constellation as this is the biggest of the three shells, and we call it Kuiper Shell 2 constellation. This constellation contains 1,296 satellites in total with 36 satellites distributed uniformly on each of the 36 orbital planes. The altitude of this constellation is 610 km and the inclination is 42°. This inclination is smaller than that of Starlink Phase 1 Version 3 constellation mentioned earlier, and thus it covers less area on the globe. The Walker constellation notation for Kuiper Shell 2 is 42°:1,296/36/$F$ and like Starlink Phase 1 Version 3, the value of $F$ for this constellation is also not defined by Amazon in its FCC filings. Fig. 2 shows the constellation pattern for Kuiper Shell 2, which is generated when using 11 as the value of $F$.

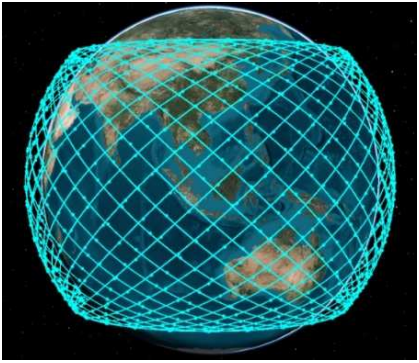

**Fig. 2.** Kuiper Shell 2 constellation.

IV. PHASING PARAMETER ANALYSIS

In this section, the concepts of phasing parameter and satellite collision avoidance as well as the methodology adopted to simulate a satellite constellation and carry out the phasing parameter analysis are discussed.

*A. Phasing Parameter F*

There are several parameters that need to be determined when designing a constellation, like total number of satellites, number of orbital planes, altitude, inclination, minimum elevation angle, eccentricity, phasing parameter, etc. The phasing parameter $F$ is an important parameter as it determines the relative phasing between satellites, and thereby the positioning of satellites in the constellation and the minimum distance between them. It can take integer values from 0 to $P$-1 with $P$ being the number of orbital planes. The relative phasing $\beta$ between satellites in adjacent orbital planes can be calculated as

$$\beta = F \times 360°/T \quad (1)$$

where $T$ is the total number of satellites in the constellation. Fig. 3 shows the relative phasing $\beta$ in a general satellite constellation. For example, in this figure, satellite 2—1 represents the first satellite on the second orbital plane. Furthermore, intra-plane phasing $\alpha$, i.e., angular spacing between satellites in the same plane, and inter-plane phasing $\gamma$, i.e., the angular spacing between adjacent orbital planes, are also shown in this figure. As the figure shows, the larger the $F$, the larger the relative phasing angle. When the phasing parameter $F$ is 0, this means that there is no relative phasing between satellites. With the increase in $F$, the relative phasing angle gets bigger, till $F$ reaches its maximum value of $P$-1.

For Starlink Phase 1 Version 3 constellation, the values of $F$ range from 0 to 21 since there are 22 orbital planes in this constellation. For Kuiper Shell 2 constellation, the value of $F$ can be taken from 0 to 35 because of 36 orbital planes in this constellation. All these values of $F$ in the two constellations need to be analyzed to find the best value, which gives the maximum value of the minimum distance between satellites in order to design the constellation for intra-constellation satellite collision avoidance. Both Starlink Phase 1 Version 3 and Kuiper Shell 2 are inclined constellations, which means that they belong to the Walker Delta type of constellations.

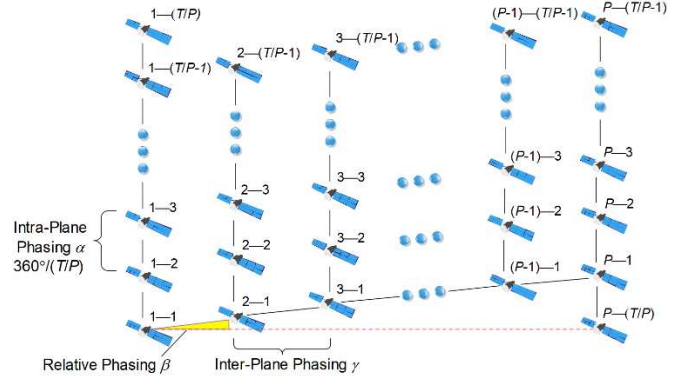

**Fig. 3.** Parameters of a satellite constellation.

*B. Satellite Collision Avoidance*

Satellites are the most important resource of a satellite constellation and satellite collision avoidance must be factored-in in the process of designing the constellation for the safety of this resource. The collision of satellites can lead to a big catastrophe, and can be even more severe, like the Kessler Syndrome. Finding a proper value of $F$ while designing the constellation will ensure the safety of satellites within the same constellation. With the change of $F$, the relative phasing between satellites in different orbital planes changes as well, and this affects the relative position of satellites in the constellation and the minimum distance between them.

To ensure collision avoidance between satellites, the size of the satellites should be taken into account. The safe distance between satellites should be several times larger than the size of satellites. Based on the FCC filings of SpaceX, a Starlink satellite is 4 m long and 1.8 m wide. It has two solar arrays and each has 6 m length and 2 m width. The size of a Kuiper satellite is not known from Amazon's FCC filings.

Based on the size of a satellite, the safe distance between the satellites in a constellation can be implied. Since the satellites are moving in space at very high speed, distance between them can shorten within very small time. The satellites within the same constellation can be regarded as having high probability of collision if the minimum distance between them is less than a certain value.

*C. Methodology for Analysis*

The satellites in a constellation should be studied based on their entire orbit around the Earth. Only in this way, we can get the complete information and data of the satellite constellation for a proper analysis. The period of a satellite in a constellation $T$ can be calculated using

$$T = 2\pi\sqrt{(r + h)^3/(GM_E)} \quad (2)$$

where $r$ is the radius of the Earth and is 6,378 km, $h$ is the altitude of the satellite, $G$ is the gravitational constant and is taken as $6.673 \times 10^{-11}$ Nm²/kg², and $M_E$ is the mass of the Earth and is taken as 5.98×10$^{24}$ kg [16]. Based on the altitude of Starlink Phase 1 Version 3 constellation and Kuiper Shell 2 constellation of 550 km and 610 km, the orbital period of a satellite in these constellations is calculated as 5,736 s and 5,810 s, respectively using (2). The satellites in these constellations take this amount of time to complete one orbit around the Earth.

Based on all the parameters of a constellation, including inclination, altitude, number of satellites and orbital planes, and phasing parameter, etc., we simulate the constellation using a satellite constellation simulator. In this study, we are interested in obtaining the minimum distance between satellites, and thus we examine the links between satellites. After defining and constructing the constellation and links between satellites, we then extract the corresponding data from the satellite constellation simulator including positions of satellites and links between satellites for further analysis via other tools like Matlab or Python.

In this work, we process this data using Python to generate more useful and systematic data, to compute the distance between satellites at each time slot. After we import the data in Python that we extract from the satellite constellation simulator, we discretize it in time. Since the orbiting satellites have very high orbital speed in space, the length of links between satellites (i.e., the distance between them) keeps changing at every second. The link data obtained from the satellite constellation simulator is continuous, and for a certain link, it contains the whole period of its existence. We discretize this data by dividing it into time slots and reorganize it so that the data shows all links that exist at a particular time slot. The position data obtained earlier is also discretized into time slots to match the discretized link data. Then based on the positions of satellites and links between them, we compute the length of each link, i.e., the distance between the pair of satellites for that link at each time slot. Next, we find the minimum distance between satellites (i.e., the link with the smallest length) at each time slot.

In this study of the phasing parameter of the two satellite constellations, minimum distances between satellites are measured at all time slots for all values of $F$ and analyzed. For each value of $F$, we select the smallest value of the minimum distance between satellites from its values at all time slots. Then, the best value of $F$ among all its possible values is the one where the value of the minimum distance between satellites is the largest. Since we are interested in finding the minimum distance between satellites, we are only interested in finding links with small distances. That's why, we check for links between satellites at every time slot for each value of $F$ that are 100 km in length or smaller. The time duration of the simulation is taken as 6,000 seconds, since this period covers the orbital period of satellites in both constellations, and it fully captures any possible close encounters between satellites within the entire orbit of the satellites. The time slot duration (i.e., the resolution of a time slot) is taken as 0.1 second. This resolution is small enough to accurately identify and capture the close encounters between satellites, i.e., occasions when they come within few meters of each other. Thus, a total of 60,000 time slots are considered to ensure the accuracy of the optimal value of the phasing parameter.

V. RESULTS

The constellations for Starlink Phase 1 Version 3 and Kuiper Shell 2 are simulated using the well-known satellite constellation simulator Systems Tool Kit (STK) version 12.1 [17]. The results for all possible values of $F$ for Starlink Phase 1 Version 3 constellation and Kuiper Shell 2 constellation are shown in Table 1 and Table 2, respectively. For each value of $F$, the value of $\beta$ calculated using (1), minimum distance between satellites in meters, the corresponding edge (or link), the time slot where the minimum distance occurs, and ranking of $F$ based on minimum distance are displayed in these tables. The column for ranking of $F$ in these tables shows the different ranks for different values of $F$ for a satellite constellation. The higher the ranking, the more suitable the value of $F$ for a constellation. For example, a ranking of 0 for $F$ is considered as the worst ranking, indicating that the satellites are very close to each other, and a high possibility of collision.

The value of $F$ which has the maximum ranking (i.e., the maximum value of the minimum distance between satellites) is highlighted in these tables in red color. In Table 1, minimum distance in Starlink Phase 1 Version 3 has the largest value when $F = 17$, and it is 61.83 km. There are many values of $F$ which are not suitable for this satellite constellation due to high possibility of intra-constellation collision, since the minimum distance is very small and is even less than 1 m, for instance, 0.57 m when $F$ equals 2. Similarly, in Table 2, the maximum value of the minimum distance for Kuiper Shell 2 is 55.89 km when $F = 11$, and this can be considered as the most suitable value of $F$ for this constellation. Also, for many values of $F$, the minimum distances are around 1 m for this constellation, which are dangerous for satellites within the same constellation due to the certainty of collision, since the size of the satellites is far greater than 1 m.

The ranking of $F$ based on the minimum distance for the two constellations is plotted in Fig. 4 and Fig. 5, and the most suitable value for $F$ can be easily seen in these figures. The values of ranking for $F$ which are not good due to a very small minimum distance are set to 0, the ranking corresponding to the largest minimum distance is the highest and the value of $F$ with highest ranking can be regarded as the most suitable value of $F$.

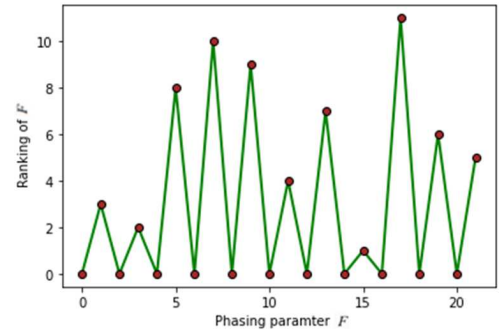

**Fig. 4.** $F$ vs. Ranking of $F$ for Starlink Phase 1 Version 3 constellation.

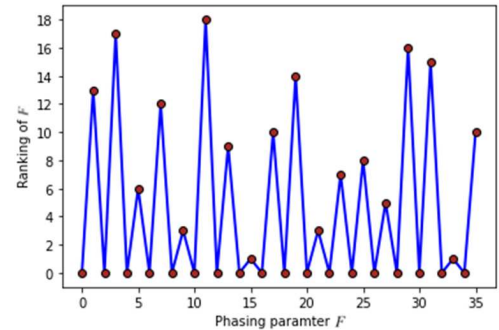

**Fig. 5.** $F$ vs. Ranking of $F$ for Kuiper Shell 2 constellation.

In Fig. 4, the best value of *F* can be clearly seen as 17 for Starlink Phase 1 Version 3 constellation since the ranking is the highest at this value of *F* among others. Other values of *F* whose ranking are bigger than 0 can also be regarded as practical values of *F*, yet 17 is the most appropriate one. Similarly, in Fig. 5 the most suitable value of *F* with highest ranking is easily visible and it is 11 for Kuiper Shell 2 constellation. According to these figures, the practical values of *F* can be taken from among half of the total possible values and these are the odd values of *F* with ranking larger than 0, while all the even values of *F* have a ranking of 0 indicating that these values are inadvisable to use due to high collision probability. Note that a similar pattern for the even values of *F* was observed in [8] during the analysis of the phasing parameter for Version 1 of Starlink's Phase 1 constellation.

In the following, we discuss two examples of a detailed analysis of the links between satellites (i.e., the distances between pairs of satellites) within a time slot. The corresponding results are shown as histograms in Figs. 6 and 7, where the x-axis represents the links that are divided into different bins (or ranges) based on their lengths, and the y-axis shows the frequency of values in different bins. The histograms in Fig. 6 and Fig. 7 are generated when *F* is 2 and 3, respectively, for Starlink Phase 1 Version 3 constellation and the time slots considered are 8328 and 4164, respectively, i.e., time slots when minimum distances occur. From these histograms, we find out that there is a small number of links between satellites with very short distances, and the majority of the links have long distances. At 8328th time slot when *F* = 2 and has an even value, 4 links exist in the 0–15 m bin while other links are in bins with distances larger than 40 km. At 4164th time slot when *F* = 3 and is odd, no link has length less than 1 km, 4 links are in the 1–2 km bin, 4 links are in the 2–5 km bin, 8 links exist in the 5–20 km bin, and 16 links exist in the 20–40 km bin, while other links have a length (i.e., distance between satellite pairs) larger than 60 km.

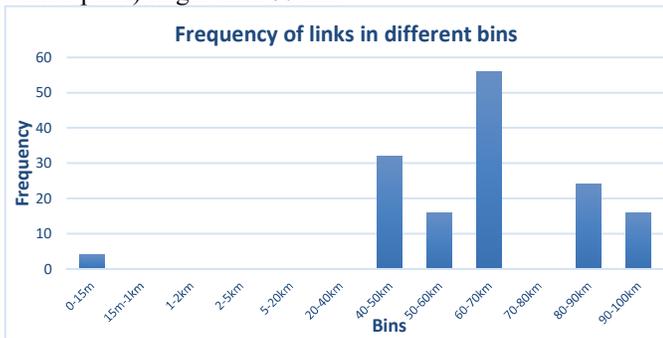

**Fig. 6.** Frequency of links in different bins for Starlink Phase 1 Version 3 constellation when *F* = 2 and time slot = 8328.

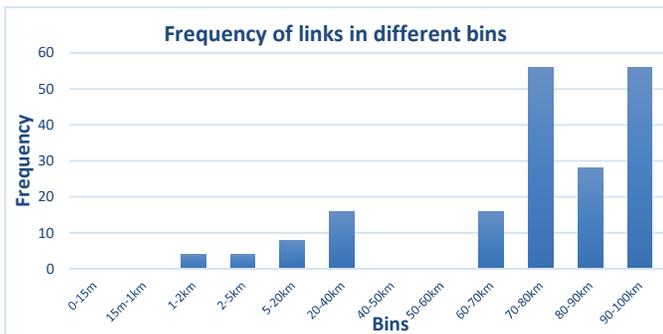

**Fig. 7.** Frequency of links in different bins for Starlink Phase 1 Version 3 constellation when *F* = 3 and time slot = 4164.

## VI. CONCLUSION

In this work, we investigate the phasing parameter for two biggest upcoming satellite constellations, namely Starlink Phase 1 Version 3 and Kuiper Shell 2. The phasing parameter determines the relative position of the satellites, which in turn affects the minimum distance between them. The satellite collisions within a constellation can be avoided if the value of *F* that provides the maximum value of the minimum distance between satellites is selected while designing a constellation. After simulating the two constellations and analyzing the minimum distance between satellites for each value of *F* in each constellation, we find the most suitable value of *F* with the highest ranking for Starlink Phase 1 Version 3 constellation as 17, and for Kuiper Shell 2 constellation as 11. Using these values of *F*, the intra-constellation collisions between satellites in these constellations can be avoided.

ACKNOWLEDGEMENT

This work has been supported by the National Research Council Canada's (NRC) High Throughput Secure Networks program (CSTIP Grant #CH-HTSN-625) within the Optical Satellite Communications Consortium Canada (OSC) framework. The authors would like to thank AGI for the STK platform.

**Table 1.** Phasing Parameter Analysis for Starlink Phase 1 Version 3 Constellation.

| $F$ | $\beta$ (°) | Minimum distance between satellites (m) | Minimum distance edge | Time slot | Ranking of $F$ |
|---|---|---|---|---|---|
| 0 | 0 | 11.74 | ('Satellite/10515', 'Satellite/11651') | 17525 | 0 |
| 1 | 0.227 | $2.32 \times 10^3$ | ('Satellite/10529', 'Satellite/11403') | 2082 | 3 |
| 2 | 0.455 | 0.57 | ('Satellite/10762', 'Satellite/11825') | 8328 | 0 |
| 3 | 0.682 | $1.80 \times 10^3$ | ('Satellite/10509', 'Satellite/10118') | 4164 | 2 |
| 4 | 0.909 | 0.57 | ('Satellite/10462', 'Satellite/11524') | 8328 | 0 |
| 5 | 1.136 | $24.50 \times 10^3$ | ('Satellite/10759', 'Satellite/11536') | 4164 | 8 |
| 6 | 1.364 | 0.57 | ('Satellite/10326', 'Satellite/11459') | 8328 | 0 |
| 7 | 1.591 | $57.96 \times 10^3$ | ('Satellite/11850', 'Satellite/12143') | 2082 | 10 |
| 8 | 1.818 | 0.57 | ('Satellite/11920', 'Satellite/10860') | 8328 | 0 |
| 9 | 2.045 | $31.86 \times 10^3$ | ('Satellite/10855', 'Satellite/11342') | 2082 | 9 |
| 10 | 2.273 | 0.57 | ('Satellite/12217', 'Satellite/11158') | 8328 | 0 |
| 11 | 2.500 | $5.00 \times 10^3$ | ('Satellite/12243', 'Satellite/12048') | 28278 | 4 |
| 12 | 2.727 | 0.57 | ('Satellite/10226', 'Satellite/11356') | 8328 | 0 |
| 13 | 2.955 | $14.98 \times 10^3$ | ('Satellite/11602', 'Satellite/11116') | 2082 | 7 |
| 14 | 3.182 | 0.57 | ('Satellite/10560', 'Satellite/11617') | 8328 | 0 |
| 15 | 3.409 | $1.63 \times 10^3$ | ('Satellite/10518', 'Satellite/11172') | 4164 | 1 |
| 16 | 3.636 | 0.57 | ('Satellite/11020', 'Satellite/12148') | 8328 | 0 |
| <span style="color:red">17</span> | <span style="color:red">3.864</span> | <span style="color:red">$61.83 \times 10^3$</span> | <span style="color:red">('Satellite/11051', 'Satellite/11536')</span> | <span style="color:red">2082</span> | <span style="color:red">11</span> |
| 18 | 4.091 | 0.57 | ('Satellite/12047', 'Satellite/10920') | 8328 | 0 |
| 19 | 4.318 | $13.27 \times 10^3$ | ('Satellite/11464', 'Satellite/10620') | 4164 | 6 |
| 20 | 4.545 | 0.57 | ('Satellite/10622', 'Satellite/11748') | 8328 | 0 |
| 21 | 4.773 | $6.12 \times 10^3$ | ('Satellite/11308', 'Satellite/11671') | 2082 | 5 |

**Table 2.** Phasing Parameter Analysis for Kuiper Shell 2 Constellation.

| $F$ | $\beta$ (°) | Minimum distance between satellites (m) | Minimum distance edge | Time slot | Ranking of $F$ |
|---|---|---|---|---|---|
| 0 | 0 | 1.12 | ('Satellite/10530', 'Satellite/12312') | 40306 | 0 |
| 1 | 0.278 | $45.07 \times 10^3$ | ('Satellite/12905', 'Satellite/13401') | 9920 | 13 |
| 2 | 0.556 | 1.11 | ('Satellite/13010', 'Satellite/11229') | 40933 | 0 |
| 3 | 0.833 | $49.49 \times 10^3$ | ('Satellite/11826', 'Satellite/13212') | 40306 | 17 |
| 4 | 1.111 | 1.12 | ('Satellite/11928', 'Satellite/10112') | 40306 | 0 |
| 5 | 1.389 | $16.58 \times 10^3$ | ('Satellite/12614', 'Satellite/10736') | 30386 | 6 |
| 6 | 1.667 | 1.12 | ('Satellite/11927', 'Satellite/10112') | 40306 | 0 |
| 7 | 1.944 | $33.93 \times 10^3$ | ('Satellite/11932', 'Satellite/13612') | 30386 | 12 |
| 8 | 2.222 | 1.12 | ('Satellite/10112', 'Satellite/11926') | 40306 | 0 |
| 9 | 2.500 | $2.82 \times 10^3$ | ('Satellite/12703', 'Satellite/13036') | 31640 | 3 |
| 10 | 2.778 | 1.11 | ('Satellite/13221', 'Satellite/11408') | 40933 | 0 |
| <span style="color:red">11</span> | <span style="color:red">3.056</span> | <span style="color:red">$55.89 \times 10^3$</span> | <span style="color:red">('Satellite/12701', 'Satellite/11318')</span> | <span style="color:red">20466</span> | <span style="color:red">18</span> |
| 12 | 3.333 | 1.12 | ('Satellite/13102', 'Satellite/11326') | 40306 | 0 |
| 13 | 3.611 | $29.14 \times 10^3$ | ('Satellite/10927', 'Satellite/11619') | 30386 | 9 |
| 14 | 3.889 | 1.11 | ('Satellite/11823', 'Satellite/13634') | 40933 | 0 |
| 15 | 4.167 | $2.66 \times 10^3$ | ('Satellite/10928', 'Satellite/12935') | 40306 | 1 |
| 16 | 4.444 | 1.12 | ('Satellite/10112', 'Satellite/11922') | 40306 | 0 |
| 17 | 4.722 | $29.46 \times 10^3$ | ('Satellite/13133', 'Satellite/12110') | 20466 | 10 |
| 18 | 5.000 | 1.12 | ('Satellite/12301', 'Satellite/10528') | 40306 | 0 |
| 19 | 5.278 | $45.42 \times 10^3$ | ('Satellite/13330', 'Satellite/11227') | 9920 | 14 |
| 20 | 5.556 | 1.12 | ('Satellite/12815', 'Satellite/11007') | 40306 | 0 |
| 21 | 5.833 | $2.82 \times 10^3$ | ('Satellite/12601', 'Satellite/12933') | 50854 | 3 |
| 22 | 6.111 | 1.11 | ('Satellite/12036', 'Satellite/10229') | 40933 | 0 |
| 23 | 6.389 | $19.93 \times 10^3$ | ('Satellite/11018', 'Satellite/11511') | 9920 | 7 |
| 24 | 6.667 | 1.12 | ('Satellite/13110', 'Satellite/11304') | 40306 | 0 |
| 25 | 6.944 | $20.73 \times 10^3$ | ('Satellite/10601', 'Satellite/13129') | 30386 | 8 |
| 26 | 7.222 | 1.11 | ('Satellite/10626', 'Satellite/12431') | 40933 | 0 |
| 27 | 7.500 | $8.41 \times 10^3$ | ('Satellite/12621', 'Satellite/12818') | 40306 | 5 |
| 28 | 7.778 | 1.12 | ('Satellite/11005', 'Satellite/12809') | 40306 | 0 |
| 29 | 8.056 | $49.21 \times 10^3$ | ('Satellite/12216', 'Satellite/13530') | 30386 | 16 |
| 30 | 8.333 | 1.12 | ('Satellite/13105', 'Satellite/11302') | 40306 | 0 |
| 31 | 8.611 | $45.42 \times 10^3$ | ('Satellite/12413', 'Satellite/10904') | 9920 | 15 |
| 32 | 8.889 | 1.12 | ('Satellite/12806', 'Satellite/11004') | 40306 | 0 |
| 33 | 9.167 | $2.66 \times 10^3$ | ('Satellite/11732', 'Satellite/13302') | 40306 | 1 |
| 34 | 9.444 | 1.11 | ('Satellite/10823', 'Satellite/12624') | 40933 | 0 |
| 35 | 9.722 | $29.46 \times 10^3$ | ('Satellite/12118', 'Satellite/13136') | 20466 | 10 |